\documentclass[article,amssymb,amsfonts,superscriptaddress,12pt]{revtex4-1}

\usepackage[utf8]{inputenc}
\usepackage[T1]{fontenc}
\usepackage{graphicx}
\usepackage{dcolumn}
\usepackage{bm}
\usepackage{xcolor}
\usepackage{subcaption}
\usepackage{amssymb}
\usepackage{amsmath}
\usepackage{epsfig}
\usepackage{siunitx}

\graphicspath{{Figures/}}

\begin{document}

\title{High repetition rate ultrashort laser cuts a path through fog}

\author{{Lorena de la Cruz}}
\affiliation{Universit\'e de Gen\`eve, GAP, Chemin de Pinchat 22, CH-1211 Geneva 4, Switzerland}

\author{{Elise Schubert}}
\affiliation{Universit\'e de Gen\`eve, GAP, Chemin de Pinchat 22, CH-1211 Geneva 4, Switzerland}

\author{Denis Mongin}
\affiliation{Universit\'e de Gen\`eve, GAP, Chemin de Pinchat 22, CH-1211 Geneva 4, Switzerland}

\author{Sandro Klingebiel}
\affiliation{TRUMPF Scientific Lasers GmbH + Co. KG, Feringastraße 10A, 85774 Unterföhring, Munich, Germany}

\author{Marcel Schultze}
\affiliation{TRUMPF Scientific Lasers GmbH + Co. KG, Feringastraße 10A, 85774 Unterföhring, Munich, Germany}

\author{Thomas Metzger}
\affiliation{TRUMPF Scientific Lasers GmbH + Co. KG, Feringastraße 10A, 85774 Unterföhring, Munich, Germany}

\author{Knut Michel}
\affiliation{TRUMPF Scientific Lasers GmbH + Co. KG, Feringastraße 10A, 85774 Unterföhring, Munich, Germany}

\author{{Jérôme Kasparian}}
\affiliation{Universit\'e de Gen\`eve, GAP, Chemin de Pinchat 22, CH-1211 Geneva 4, Switzerland}
\email{jerome.kasparian@unige.ch}

\author{Jean-Pierre Wolf}
\affiliation{Universit\'e de Gen\`eve, GAP, Chemin de Pinchat 22, CH-1211 Geneva 4, Switzerland}

\date{\today}

\begin{abstract}
We experimentally demonstrate that the transmission of a 1030~nm, 1.3~ps laser beam of 100 mJ energy through fog increases when its repetition rate increases to the kHz range. 
Due to the efficient energy deposition by the laser filaments in the air, a shockwave ejects the fog droplets from a substantial volume of the beam, at a moderate energy cost. This process opens prospects for applications requiring the transmission of laser beams through fogs and clouds.
\end{abstract}

\maketitle

\section{Introduction}
Lasers offer many actual or prospective applications at the atmospheric scale~, including~\cite{KaspaW2008}, the remote delivery of high-intensity for surface ablation \cite{Garcia2000, StelmRMYSKAWW2004, FujiiGMNN2006, TzortAG2006}, free-space communications~\cite{Kiasaleh2006,Belmonte1997,PolynPKRM2007}, remote sensing \cite{KaspaRMYSWBFAMSWW2003,GalveFIMIY2002,Gravel2004,Hemmer2011,Svanb},  or weather modulation \cite{KaspaAAMMPRSSYMSWW2008a,HeninPRSHNVPSKWWW2011,Ju2014}.
In that purpose, efficient transmission of the beam through the atmosphere is essential, and cloud and fogs constitute obvious obstacles. 

When the incident peak power of a laser pulse exceeds a critical power ($P_\textrm{cr} = 5$~GW in air at 1030~nm), focusing and defocusing non-linearities including  the laser-generated plasma and the saturation of the medium polarisability under strong-field illumination~\cite{BejotKHLVHFLW2010a,Volkova2011,Richter2013} result in a self-guided propagation regime:
Filamentation \cite{BraunKLDSM1995,ChinHLLTABKKS2005,CouaiM2007,BergeSNKW2007}.
 
Laser filaments have a diameter of \SIrange{100}{200}{\micro\meter} and are surrounded by a "photon bath" carrying most of the beam energy. This energy is able to re-create them after they have been blocked by an obstacle like a water drop~\cite{CourvBKSMYW2003,KolesM2004,SkupiBPL2004}.  However, since the photon bath undergoes elastic losses, the aerosols ultimately limit the filamentation length. Indeed, a filament cannot propagate without being fed by its photon bath~\cite{LiuTAGBC2005}. 

Laser filaments leave behind them a cylindric region where the air density is depleted~\cite{Thermal1} (a  ``density hole") with a lifetime of hundreds of microseconds~\cite{Thermal2,Thermal3,Cheng2013}. In this work, we show that the associated shockwave "cleans" the atmosphere not only in the filament, but also a significant fraction of the photon bath. 

This approach substantially differs from attempts to clear clouds and fogs in the 70’s and ‘80s with high-energy CO$_2$ lasers, in which prohibitively high intensities and energies are needed to evaporate and shatter water drops (typically 10~kW/cm$^2$ continuous wave lasers~\cite{Zuev1984} and 1--1000~MW/cm$^2$ pulsed lasers~\cite{Kwoh1988,Pustovalov1992}, respectively). Conversely, the energy of the shockwave can expell particles out of the beam, at a much lower energy cost. 

By investigating the propagation of ultrashort (ps) laser pulses in the near-infrared through a dense fog, we show that the transmission increases with the repetition rate, i.e., with the average beam power. This increased transmission is due to the above-mentioned shockwave, expelling the particles from the central region of the beam, well beyond the volume of the filament itself. This thermo-mechanical effect is efficient beyond the filament volume and offers a prospect to improve laser beam transmission through fog and clouds.

\section{Experimental setup}

\begin{figure} 
	\centering
		\includegraphics[width=0.70\columnwidth]{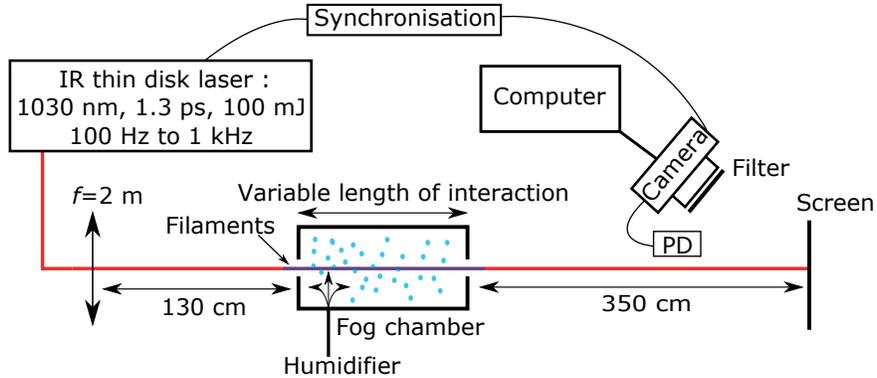}
	\caption{Experimental setup.}
	\label{fig:montage}
\end{figure}

As sketched in Figure \ref{fig:montage}, the experiment consisted in
propagating the beam of an ultrashort laser beam through a chamber filled with fog.\\
The laser system was an Yb:YAG thin disk laser (Dira, TRUMPF Scientific Lasers GmbH + Co. KG~\cite{Klingebiel2015}), delivering 1.3~ps pulses with 100~mJ energy at a wavelength of 1030~nm. The beam was slightly focused ($f$~=~2~m). The repetition rate was varied between 100 and 1000 Hz, corresponding to average powers of 10 to 100 W. The peak power of 75~GW corresponds to 15 critical powers. As a result, typically 3--4 filaments of 50 cm length were observed on the screen. Considering that each of them carries an energy $E_\text{fil}$~=~5~mJ, 3 filaments carry 15~mJ, i.e, 15\% of the total beam energy of 100~mJ.

3.5~m after the fog chamber, the beam was imaged on a screen and its profile recorded by a PixelInk PL-B761U CCD camera with 480 x 752~pixels. Approximately 1900 single-shot images were recorded for each experimental condition through a Schott BG7 filter, that blocked the continuum in the 700--900~nm spectral region, but only attenuates the fundamental wavelength. 
The energy transmitted through the fog was calculated as the profile-integrated fluence on the screen, and normalized by a reference measurement without fog.

The fog was produced by a large volume droplet generator and introduced into a 40~cm long chamber. Leaks through the openings implied an interaction length of 50~cm between the laser beam and the fog. Alternatively, the fog at the exhaust of the generator was directly blown onto the laser beam, offering propagation through 6~cm of fog. In either case, the fog interacted with the most intense longitudinal section of the filament, i.e., where the plasma noise was strongest.

\begin{figure}[b]
	\centering
	\includegraphics[width=0.70\columnwidth]{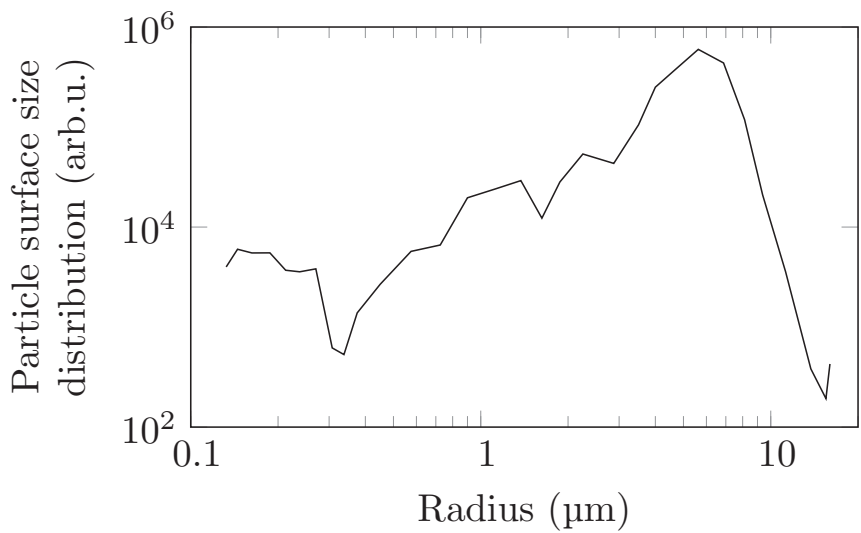}
	\caption{Particle surface size distribution measured by the Grimm 1.109 in the fog chamber.}
	\label{fig:Grimm}
\end{figure}

The fog droplet size distribution (Figure~\ref{fig:Grimm}) 
was measured by using an optical aerosol sizer (Grimm, model 1.109). We checked that the size distribution was homogeneous across the fog chamber, and cross-checked the typical size with direct imaging of the particles impacted on a glass surface. 
The mode of this size distribution is $a_\mathrm{eff}=\SI{5}{\micro\meter}$, compatible with typical fog conditions in the atmosphere. In order to reach optical densities encountered in fogs on the meter-scale of the laboratory, the concentration of droplets was increased by a factor of typically 100 as compared with actual fogs.

\section{Results and discussion}

\begin{figure} [t]
	\centering
	\includegraphics[width=0.70\columnwidth]{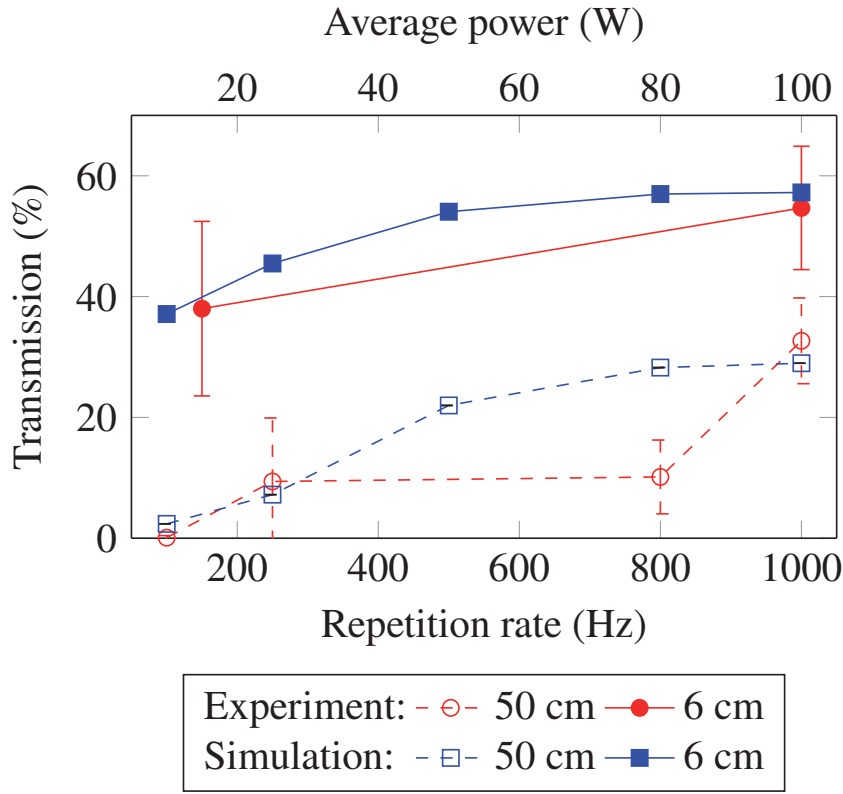}
	\caption{Beam transmission of 1030~nm, 100~mJ, 1.3~ps laser pulses through the fog chamber for different interaction lengths and repetition rates.}
	\label{fig:perte_IR}
\end{figure}

Figure~\ref{fig:perte_IR} displays the beam transmission as a function of the repetition rate of the laser. Higher repetition rates, associated with average beam powers up to 100~W, clearly increase the transmission of the beam through the fog, from 0.1\% at 100~Hz, to 32\% at 1~kHz. As mentioned above, the three filaments carry only 15\% of the beam energy. The observed increase in the transmission therefore implies that at high repetition rates the laser transmission also increases in the lower-intensity photon bath.

  \begin{figure} [t!]	
  	\centering
	\includegraphics[width=0.70\columnwidth]{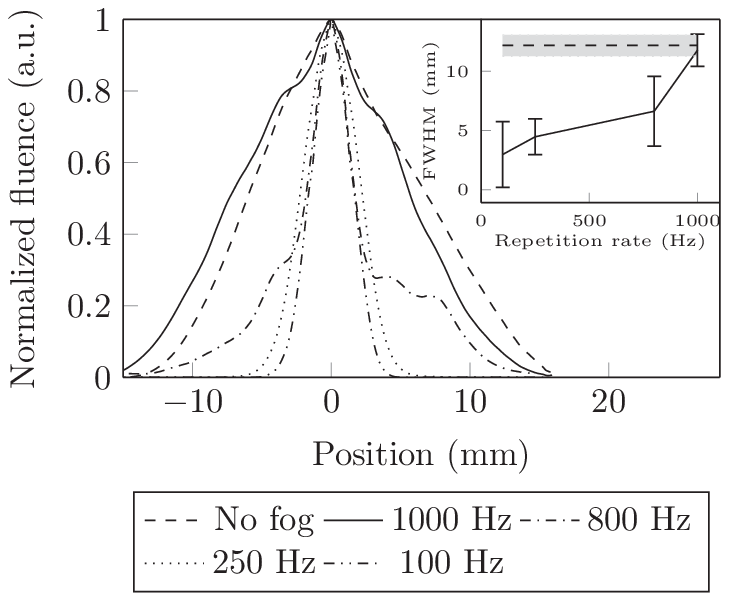} 
	  	\caption{Typical fluence profile of the beam on a screen after propagating through the fog, as a function of the repetition rate. Inset: Corresponding beam diameter (FWHM). The dotted line indicates the beam diameter without fog, the gray region marking the error bar.}
  	\label{profil}
  \end{figure}

Indeed, the higher transmission is associated with an wider transmitted beam (Figure~\ref{profil}). While at 100~Hz repetition rate only the central part of the profile on the screen is (partially) transmitted, at 1~kHz this part covers almost the whole beam.
In other words, increasing the repetition rate (hence, the average power and the deposited heat) increases both the transmission in the beam center and the width of the region over which this transmission is increased.

The vaporization or the shattering of the water droplets present in the filament volume would at most result in the full transmission of the filaments, that as discussed above carry only 15\% of the beam energy. It cannot further increase the beam transmission, nor extend the size of the transmitted region of the beam. 
To explain these observations, droplets have to be ejected also from the photon bath. This is made possible by the shockwave in the air associated to local heating~\cite{Thermal3}, due to the cumulative energy deposition~\cite{Thermal2} by the filaments in the air (typically 2\% at 1~kHz~\cite{Houard2016a}).

The heat deposited by filaments in conditions comparable to ours typically generates a depleted channel with a local pressure reduced to $P\approx0.5$~atm~\cite{Thermal1,Thermal3},
implying that the corresponding air parcel doubles volume. As a consequence, we assumed that the shockwave sweeps a volume at least twice as large as that of the filaments, i.e., a cylinder of \SI{100}{\micro\meter} radius for a filament radius of \SI{70}{\micro\meter}.
Based on the temporal evolution of the air density calculated by ~\cite{Thermal1}, we considered an air radial expansion speed $v_\mathrm{air} =21$~m/s for $\SI{0}{\micro\second} \le t \le \SI{1}{\micro\second}$ and a collapse speed $v_\mathrm{air} = -10$~mm/s for $\SI{100}{\micro\second} \le t \le \SI{2}{\milli\second}$.
Outside of these two time intervals, we consider that the air stays still.
We solved the equation of motion of the droplets under aerodynamic drag~\cite{Frottement}, assuming that they keep their spherical shape:
\begin{equation}  m_\text{drop} \cdot \frac{\textrm{d} \vec{v}_\text{drop}}{\textrm{d}t}= \frac{1}{2} \pi a^2 \rho_\text{air} C_d  \left( \vec{v}_\text{air} -\vec{v}_\text{drop}  \right) ^2 
\label{drag}
\end{equation} 
where $\rho_\text{air}$ is air density and $C_d= 0.47$ is the drag coefficient for a sphere.
We find that the shockwave has a net effect of ejecting the droplets of radius $a = a_\textrm{eff}=\SI{5}{\micro\meter}$ at a speed of $v_\textrm{out} = 60$~mm/s out of both the filament and its surrounding photon bath.

The balance between the particle ejection by the shockwave and the advection by the transverse wind can be characterized by calculating the net flux $\phi$ of particles through the outer surface of the region swept by the shockwave, modeled as a cylinder of length $L$ and radius $R=\SI{100}{\micro\meter}$:

\begin{equation}
\phi_\text{net}=\frac{\textrm{d}N}{\textrm{d}t}=n_\text{drop}v_\text{in} 2 R L - \frac{2N(t)v_\text{out}}{R}
\label{eq:balance}
\end{equation}
where $N(t)$ is the number of particles in the considered volume, $t$ is the time, and $n_\text{drop}$ is the drop concentration far from the filament (as supplied by the droplet generator). The advection speed $v_\text{in}$ of the incoming droplets is estimated to be 12~mm/s based on the droplet generator flow and chamber geometry.

 \begin{figure} 
 \centering
 	\includegraphics[width=0.70\columnwidth]{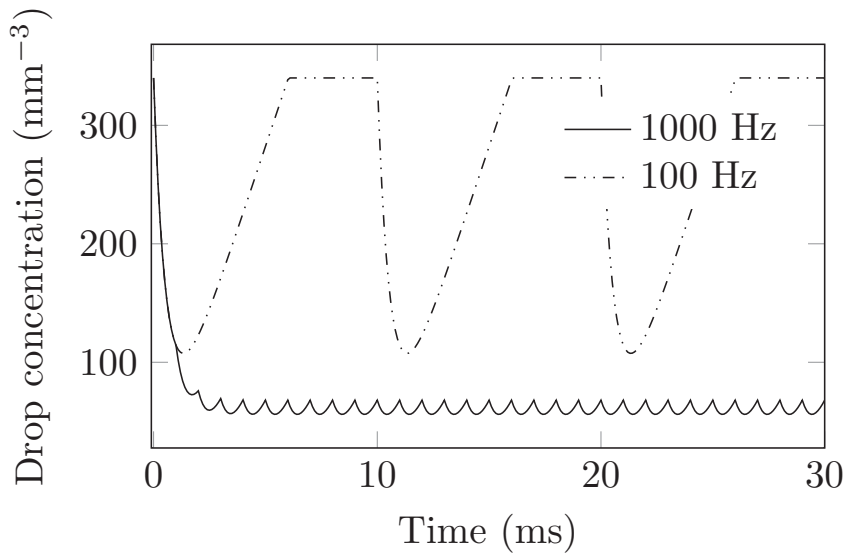}
 	\caption{Evolution of the average drop concentration in the \SI{100}{\micro\meter} cylinder swept by the shockwave}
 	\label{fig:drop}
 \end{figure}

Figure~\ref{fig:drop} displays the droplet concentration evolution within the cylinder swept by the shockwave, based on integrating Eq.~(\ref{eq:balance}). At 100~Hz, the advection is sufficiently fast to replenish the droplet concentration between two pulses. In contrast, at 1~kHz, the advection is too slow, so that the droplet concentration decreases. After 3--5 pulses, it  reaches a steady regime  in which advection and droplet expulsion by the shockwave balance each other. The droplet concentration has then dropped by a factor of almost 6 from 344 to 60~mm$^{-3}$. The  concentration estimated by this approach is independent from the diameter of the considered cylinder, because the drag force of the air is insufficient to slow them down significantly over a few milliseconds: Their speed can be considered constant at the investigated time scales.

To precisely estimate the effect of droplet ejection by the shockwave on the transmission of the beam, we modeled in two dimensions the interaction of  droplets carried by an advection flow, under the influence of the laser-generated shockwave. More specifically, we performed Monte-Carlo simulations of the trajectory of \num{40000} droplets of radius $a_\textrm{eff}=\SI{5}{\micro\meter}$, with an initial  advection speed $v_\textrm{in}$=12~mm/s transverse to the laser beam. 
At each laser pulse, droplets located within the area swept by the shockwave associated with each of the three filament are ejected radially at a speed of 60~mm/s.
The resulting droplet concentration map is then used to compute Mie scattering for the effective particle radius and hence determine the local transmission, that is then integrated over the beam fluence profile to yield the beam transmission value. 
Propagation lengths of both 50~cm and 6~cm, representative of the experiments, have been considered.

In order to match the experimental parameters, the beam is modeled as a Gaussian beam with a FWHM diameter of 5~mm, with 3 filaments evenly positioned on a circle with 0.7~mm radius around the beam center. 
The orientation of the triangle formed by the 3 filaments evolves randomly from shot to shot. We checked that the results were insensitive to the details of this motion.

A droplet-depleted region appears downstream of the filamenting region.
The corresponding rise of the transmission of the photon bath contributes to one half of the the observed transmission increase at high repetition rates. As shown in Figure~\ref{fig:perte_IR}, this model reproduces well the experimental results for an initial droplet concentration of 344~mm$^{-3}$, demonstrating the key role of the shockwave in clearing the fog and increasing the laser beam transmission.
Note that the discrepancy with the experimental point at 800~Hz is due to a lower stability of the laser beam at this repetition rate, that limits the cumulative thermal effect.
Although the size mode of actual fog is slightly lower ($\sim \SI{1}{\micro\meter}$), one can expect that our results can be extended at least to atmospheric ranges in a dense fog.

\section{Conclusion}
As a conclusion, we investigated the transmission of a high-average power, high-repetition rate ultrashort laser beam through a dense fog. Due to the energy deposition in the air, a shockwave expels the fog droplets not only within the filament volume, but also from the photon bath. As a consequence, a drastically improved beam transmission is observed. This effect increases for higher repetition rates where the balance between the said expulsion and the advection of new background droplets favors more the former. 

This work therefore opens  prospects to improved laser beam transmission through fog, opening promising perspectives to point-to-point laser communication, remote sensing, or lightning control through clouds.

\textbf{Acknowledgments}
We acknowledge financial support from the ERC advanced grant « Filatmo ».
The authors thank M. Moret for technical support.

\bibliography{Biblio2}

\end{document}